\documentclass[twocolumn,preprintnumbers,amsmath,amssymb]{revtex4}

\usepackage{amsmath}
\usepackage{amssymb}
\usepackage{graphicx}
\usepackage{dcolumn}
\usepackage{bm}
\usepackage{ulem}

\newcommand{\beq}{\begin{equation}}
\newcommand{\eeq}{\end{equation}}
\newcommand{\bea}{\begin{eqnarray}}
\newcommand{\eea}{\end{eqnarray}}
\newcommand{\Zav}{\langle Z \rangle} 
\newcommand{\Uav}{\langle U \rangle} 
\newcommand{\Dp}{{\Delta}^+} %
\newcommand{\Dm}{{\Delta}^-} %
\newcommand{\Dpk}{\Delta^+_{\bm{k}}} %
\begin{document}

\title{Short-range ordering and equilibrium structure of 
binary crystal mixtures of atomic nuclei in white dwarf cores}
\author{D. A.\ Baiko}

\affiliation{Ioffe Institute, Politekhnicheskaya 26, 194021 Saint Petersburg, Russia}

\date{\today}

%
%
%
%
%
%
%
\begin{abstract}
Mixtures of bare atomic nuclei on a nearly uniform degenerate electron background
are a realistic model of matter in the interior of white dwarfs.
Despite tremendous progress
in understanding their phase diagrams achieved mainly via
first-principle simulations, structural, thermodynamic, and kinetic
properties of such mixtures are poorly understood. 
We develop a semi-analytic model of the crystal state of binary mixtures
based on the concept of mutual short-range ordering of ions of different sorts.
We derive analytic formulas for electrostatic energy
of crystal mixtures, including the effect of static ion displacements from the 
lattice nodes, and 
estimate their residual entropy. Then we perform free
energy minimization with respect to the order parameters for a C/O mixture at all 
relevant compositions and temperatures.
The resulting C/O phase diagram is in a reasonable agreement with 
that obtained in the most recent first-principle study. 
The equilibrium microstructure of a crystallized
mixture is shown to evolve with decrease of temperature which, in principle,
can induce structural transitions. The latter will be accompanied
by thermal energy release.     
The proposed theory opens up a path to analyze ordering and construct phase diagrams 
of ternary mixtures, 
which are of great practical interest in astrophysics,
as well as 
to improve calcuations of electron-ion scattering rates and 
kinetic properties of dense crystallized matter.
\end{abstract}


\maketitle




{\it Introduction} -- Matter in the inner layers of white dwarfs (WD) 
and in the outer neutron star crust is 
compressed by immense gravity of these objects to densities which may be 
as high as $\sim 10^{10}$-$10^{11}$ g/cm$^3$. The result is a complete stripping
of electrons off the atomic nuclei and a formation of ion plasma with 
a nearly uniform background of degenerate electrons. The temperature of matter
quickly drops below typical ion interaction energies and the ions
condense into a strongly coupled liquid which later freezes into a solid. 
In principle, nuclei of several different sorts may be present simultaneously
at a given density. Physical properties of such ion mixtures
are important for various applications, particularly, for stellar evolution modeling. 
For instance, it has been shown recently that
$^{22}$Ne distillation controlled by the phase diagram of a crystallizing 
ternary ion mixture may cause multi-billion year cooling delays of WD 
\cite{I+91,BDS21,BBC24}. 

Understanding phase diagrams of ion mixtures has been steadily improving
thanks to impressive research over the past several decades 
\cite{S76,HTV77,S80,BHM88,IIO88,SC93,DSC96,S96,DS03}.
The decisive progress has finally been achieved through
advanced first-principle simulations 
\cite{O+93,MC10,HSB10,H+12,CHC20,BD21,BD21b,C+21,CBF23,BDS21}. 
While simulations of classic liquid mixtures, 
especially at moderate couplings, are seemingly straightforward, this is 
not so for the crystal state
where there are mutually conflicting requirements. One needs higher $N$
to be closer to the thermodynamic limit but, on the other hand,
crystal simulations tend to stick to an initial ion arrangement and thus
one has to explicitly explore ion permutations, possible number of which rapidly grows with $N$.
As a result, the actual structure of multicomponent crystal mixtures
is currently unknown to the point that it is unclear whether placement
of different ions onto lattice nodes should be treated as random or not.

In this work, we investigate,
for the first time,
the microstructure of binary crystal mixtures
of atomic nuclei. 
We do that semi-analytically
by introducing order parameters, describing average relative positions 
of ions of two sorts in a lattice, by expressing electrostatic energy and residual entropy via these parameters,
and by minimizing free energy with respect to them. In this way, we establish that
the crystal mixture is not fully disordered but prefers a short-range order
which, under the assumption of thermal equilibrium, evolves with temperature decrease.

{\it Order parameters and electrostatic energy} -- Suppose that we have 
a binary mixture of ions with charge numbers
$Z_1$ and $Z_2$, $Z_1<Z_2$, and charge-neutralizing rigid background of
electrons with density $n_{\rm e}$. Let the number fraction
of $Z_2$ ions be $x$. The average ion charge number is then
$\Zav = x Z_2 + (1-x) Z_1$. Suppose that the ions are arranged in
an ideal crystal lattice, one of the nodes of which is chosen as the origin
while all the others are specified by lattice vectors $\bm{R}$. The 
number density of the lattice nodes is $n_{\rm i}=n_{\rm e}/\Zav$. If the 
lattice is assumed to be of the bcc type, its spacing $a_{\rm l}=(2/n_{\rm i})^{1/3}$.

The structure of the mixture is fully determined by specifying the function
$Z(\bm{R})$ which can take values $Z_1$ and $Z_2$. Let us decompose this
function as 
$Z(\bm{R}) = \Zav + \delta Z(\bm{R}) = \Zav + Z^+ \Dp(\bm{R}) - Z^- \Dm(\bm{R})$,
where $\Dp(\bm{R})=1$ for nodes
occupied by $Z_2$-ions and 0 otherwise, $\Dm(\bm{R})=1-\Dp(\bm{R})$
(see also \cite{K24b}).

Let us consider the expression
\beq
   P^+(\bm{R}) = \frac{1}{N} \sum_{\bm{R}_1} \Dp(\bm{R}_1) \Dp(\bm{R}_1+\bm{R})~,
\label{P+}
\eeq
where $N$ is the total number of ions and thermodynamic limit
is understood. $P^+(0) = x$. If lattice vectors $\bm{R}_{\rm 1n}$ and 
$\bm{R}_{\rm 2n}$ point to a nearest 
neighbor and to a next nearest neightbor, which are always occupied, say,
by a different ion and by an ion of the same sort, respectively, then 
$P^+(\bm{R}_{\rm 1n})=0$ and $P^+(\bm{R}_{\rm 2n})=x$. If, at large $\bm{R}$,
the mutual placement of ions becomes independent, $P^+(\bm{R}\to \infty) \to x^2$.
If the limit is different from $x^2$ or if there is no limit at all
(e.g. in a perfectly ordered CsCl crystal at $x=1/2$, $P^+$ oscillates 
between 0 and 1/2), there is a long-range order.    

In this work, we shall assume that there is no long-range order (but see \cite{K24}). At finite
arguments, $P^+(\bm{R}_{i{\rm n}})$ is isotropic, i.e. the same
for all neighbors of the same order, and we denote it as $\alpha_i$, the $i$th 
short-range order parameter. We emphasize that the definition
(\ref{P+}) implies averaging over $\bm{R}_1$, which means that the actual
composition in the vicinity of any particular ion can be essentially arbitrary.
The concept of short-range and long-range order is not new:
see, e.g., \cite{C65} and references therein.
However, it has never been applied to dense mixtures of bare atomic nuclei. 
Our definition of $\alpha_i$ is slightly different from that of \cite{C65}. 

In Table \ref{neighbors},
we collect several parameters for the first 8 shells of nearest neighbors
in the bcc lattice. 
In particular, we show cartesian coordinates of a vector, belonging to each shell, from
which coordinates of all the other equivalent vectors can be obtained by symmetry
transformations (arbitrary combinations of coordinate permutations and 
sign changes), the vector length, $l$, the total number of neighbors in the $i$th shell, $m_i$.

Expanding $\Dp(\bm{R})$ in the Fourier-integral  
\beq
   \Dp(\bm{R}) = \int_{{\rm B}_1} \frac{{\rm d}\bm{k}}{(2\pi)^3 n_{\rm i}}
   \left[x (2\pi)^3 n_{\rm i} \delta(\bm{k}) + \Dpk \right] \, {\rm e}^{i\bm{k}\bm{R}}~,    
\label{fur_r}
\eeq
where $\Dpk$ is regular and $\bm{k}$ is any vector in the first Brillouin 
zone, ${\rm B}_1$, 
one can deduce that 
\beq
   \frac{1}{N} \left| \Dpk \right|^2 = x(1-x) 
   + \sum_{i=1}^{\infty} (\alpha_i - x^2) \sum_{\bm{R}_{i{\rm n}}} \cos{(\bm{k}\bm{R}_{i{\rm n}})}~.
\label{Dpkres}
\eeq
Naturally, the left-hand side of equation (\ref{Dpkres})
must be non-negative at all $\bm{k}$ which restricts
allowed values of $\alpha_i$.     

On average for a $Z_2$-ion, $m_i \alpha_i/x$ neighbors of the 
$i$th order are also $Z_2$-ions, whereas for a $Z_1$-ion, 
$m_i (1-2x + \alpha_i)/(1-x)$ neighbors of the 
$i$th order are $Z_1$-ions. The total number of these ions should be 
non-negative and $\leq m_i$ as well as the same
for a short-range ordered and for a completely disordered system.
This produces further constraints: $\max{(0,2x-1)} \leq \alpha_i \leq x$
and
\beq
    \sum_{i} m_i (\alpha_i-x^2) = 0~.
\label{sum_rule}
\eeq

With these definitions, the electrostatic energy per ion, assuming that 
all ions are situated exactly at their lattice nodes, 
can be written as
\beq
   U = \Uav + U_{\rm corr} = -\zeta \frac{\Zav^2 e^2}{a_{\rm i}} + 
   \frac{e^2}{2N} \sideset{}{'}\sum_{\bm{R},\bm{R}_1} 
   \frac{\delta Z(\bm{R}) \delta Z(\bm{R}_1)}{|\bm{R}-\bm{R}_1|}~, 
\nonumber
\eeq
where $\zeta$ is the Madelung constant and $a_{\rm i} = (4\pi n_{\rm i}/3)^{-1/3}$.
Note that $\Uav$ is different from the standard linear mixing formula (e.g., \cite{S76}), which contains
$\langle Z^{5/3} \rangle$ instead.  

The second term reduces to 
\beq
    U_{\rm corr} = \frac{(Z_2 - Z_1)^2 e^2}{2} 
    \sum_{i=1}^{\infty} (\alpha_i - x^2) I_i~,   
\label{Ucorr_res}
\eeq
where $I_i = m_i/R_{i{\rm n}}$. 
If the ion placement is completely random, 
$\alpha_i = x^2$ for any $i$, $\left| \Dpk \right|^2 = N x(1-x)$ throughout ${\rm B}_1$,
whereas $U_{\rm corr} =0$.

\begin{table}
\begin{center}
\begin{tabular}{cccccc}
\hline
\hline
 $i$   & $\bm{R}_{i{\rm n}}$  &  $4l^2/a^2_{\rm l}$ & $m_i$  & $a_{\rm i} T_i$ & $a_{\rm i}(I_i - T_i)$  \\
\hline
 1   & $a_{\rm l} (1/2,1/2,1/2)$ & 3 & 8 &  4.4334 & 0.1149 \\   
\hline 
 2   & $a_{\rm l} (1,0,0)$ & 4 & 6 &  3.0577 & -0.1035 \\  
\hline 
 3   & $a_{\rm l} (1,1,0)$ & 8 & 12 &  4.1411 & 0.0368 \\   
\hline 
 4   & $a_{\rm l} (3/2,1/2,1/2)$ & 11 & 24 &  7.2268 & -0.1009 \\  
\hline 
 5   & $a_{\rm l} (1,1,1)$ & 12 & 8 &  2.1968 & 0.0774 \\   
\hline 
 6   & $a_{\rm l} (2,0,0)$ & 16 & 6 &  1.4950 & -0.0179 \\  
\hline 
 7   & $a_{\rm l} (3/2,3/2,1/2)$ & 19 & 24 &  5.3710 & 0.0510 \\   
\hline 
 8   & $a_{\rm l} (2,1,0)$ & 20 & 24 &  5.2993 & -0.0146 \\  
\hline
 0  &&&& 0.90183 & \\
\hline
\hline
\end{tabular}
\end{center}
\caption{Neighbors of the first 8 orders and numerical values of $T_i$ and $I_i-T_i$}
\label{neighbors}
\end{table}

Static displacements of ions ($\ll a_{\rm l}$) from the lattice nodes 
under the action
of local forces not completely compensated due to lack of symmetry,
may be expected to give rise to an energy contribution, $U_{\rm shft}$, 
of the same order of magnitude as $U_{\rm corr}$. After some algebra, it can be shown that 
to the lowest, $(\delta Z)^2$, order
\beq
    U_{\rm shft} = - \frac{(Z_2-Z_1)^2 e^2}{2} [ x (1-x) \, T_0 +
      \sum_{i=1}^{\infty} (\alpha_i - x^2)\, T_i ]~.
\label{Ushft} 
\eeq
In this case, 
\beq
    T_i = \int_{{\rm B}_1} \frac{{\rm d}\bm{k}}{(2\pi)^3 n_{\rm i}} 
    \sum_{s=1,2,3} \frac{\left|\bm{e}_{\bm{k}s} \cdot \bm{b}_{\bm{k}}\right|^2}{\lambda_{\bm{k}s}}
     \sum_{\bm{R}_{i{\rm n}}} \cos{(\bm{k}\bm{R}_{i{\rm n}})}
\label{Tdef}
\eeq
at $i>0$. $T_0$ has 1 in place of $\sum_{\bm{R}_{i{\rm n}}}$, whereas  
$\lambda_{\bm{k}s}$ are eigennumbers of
a real symmetric 3$\times$3 matrix $D_{\bm{k}\alpha \beta}$,
and $\bm{e}_{\bm{k}s}$ are its three respective mutually orthogonal unit
eigenvectors. Finally,  
\bea
     D_{\bm{k}\alpha \beta} &=& \sum_{\bm{R}\ne 0} \left( {\rm e}^{i\bm{k}\bm{R}} -1 \right)
     g_{\alpha\beta}(\bm{R}) + n_{\rm i} \int {\rm d}\bm{r} \, g_{\alpha\beta}(\bm{r})~,    
\label{Ddef} \\
     \bm{b}_{\bm{k}} &=& - \sum_{\bm{R}\ne 0} {\rm e}^{i\bm{k}\bm{R}}
     \, \bm{h}(\bm{R})~,
\label{bdef}             
\eea
where $g_{\alpha\beta}(\bm{r}) \equiv \delta_{\alpha\beta}/r^3 - 3r_\alpha r_\beta/r^5$
and $\bm{h}(\bm{r}) \equiv \bm{r}/r^3$. 
Practical expressions for the matrix
$D_{\bm{k}\alpha \beta}$ and the vector $\bm{b}_{\bm{k}}$ can be obtained with
the standard Ewald method. At low $k$,
$\bm{b}_{\bm{k}} \approx 4 \pi i n_{\rm i} \bm{k}/k^2$.   

\begin{figure}
\begin{center}
\includegraphics[bb=26 25 611 563, width=88mm]{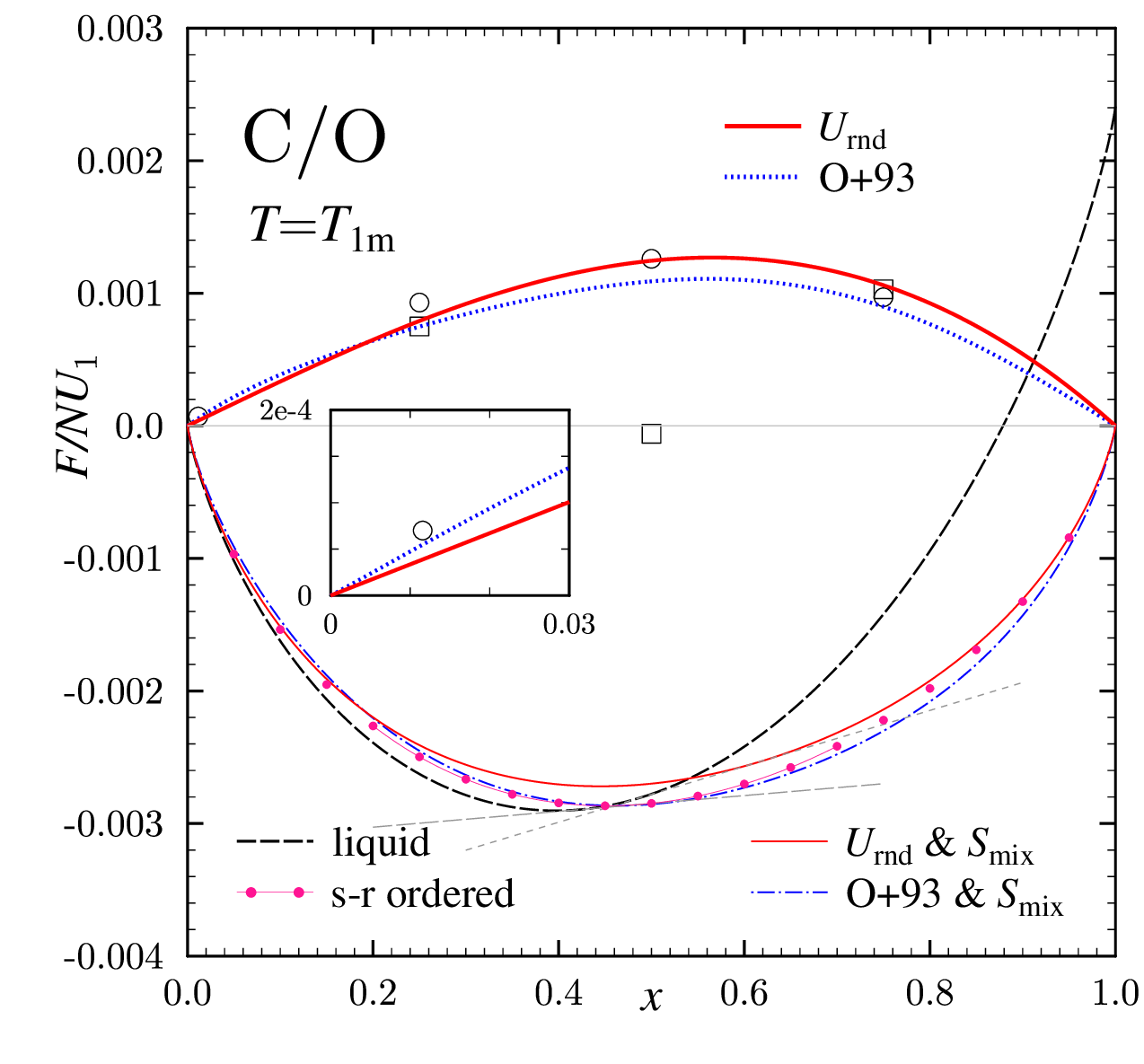}
\end{center}
\vspace{-0.2cm}
\caption[ ]{Various energy and free energy models for C/O mixture at 
$T=T_{\rm 1m}$. Corrections to linear-mixing energies of the crystal 
reported in Ref.\ \cite{O+93}: ``minimum'' MC energies (open circles), 
energies of fully-ordered configurations (squares), fit (dense dotted). 
Thick solid curve shows $U_{\rm rnd}$. Helmholtz free energies: based on 
the fit \cite{O+93} (dot-dashed), $U_{\rm rnd}-TS_{\rm mix}/N$ (solid), 
minimized with respect to the short-range order parameters 
(filled dots with a thin solid line, going through them), of the 
liquid (black dashes). Straight grey dashes illustrate double tangents.} 
\label{dfs}
\end{figure}

We have calculated the integrals $T_0$-$T_8$ on a dense grid
of wavevectors in ${\rm B}_1$ and the results are given in Table \ref{neighbors}.  
Their inaccuracy can be estimated as $\pm 1$ in the last digit. 
The degree of compensation between seemingly unrelated $T_i$ and $I_i$ is 
quite remarkable. $U_{\rm shft}$, unlike $U_{\rm corr}$, is non-zero for a completely disordered
binary mixture. Its electrostatic energy thus is
$U_{\rm rnd}=\Uav +U_{\rm shft} = -e^2 [\zeta \Zav^2/a_{\rm i} 
+ 0.5 \, T_0 \, (Z_2-Z_1)^2 x (1-x)]$ (cf.\ Fig.\ \ref{dfs}).

For structural studies, one also requires residual entropy of the static lattice,
i.e. the entropy which remains in the crystal mixture at zero temperature.
For a completely disordered binary solid, the results is
\beq
  S_{\rm mix} = - Nx \ln{x} - N(1-x) \ln{(1-x)}~. 
\label{Smix}
\eeq
Ref.\ \cite{O+93} used instead a slightly different quantity 
\beq
   S_{\rm mixZ} = S_{\rm mix} + N(\ln{\Zav}-\langle \ln{Z} \rangle)~,
\label{SmixZ}
\eeq
which we shall also utilize further,
however, as discussed in \cite{B23}, the choice of $S_{\rm mixZ}$ 
was not convincingly justified. On the other hand, the residual entropy
of a perfectly ordered crystal mixture, such as the CsCl system, is 0.
It is then natural to expect that a crystal mixture with the short-range order
considered in the present work would have a residual entropy intermediate
between 0 and $S_{\rm mix}$. We adopt an entropy estimate based on the idea of \cite{C65}. The details 
are provided in the Appendix.

{\it Random C/O mixture} -- Let us apply the above theory to a dense 
C/O mixture. We shall plot 
energies and Helmholtz free energies with the linear mixing electrostatic energy,  
$-\zeta \langle Z^{5/3} \rangle e^2/a_{\rm e}$, subtracted, and in units of
$U_1 \equiv Z_1^{5/3} e^2/a_{\rm e}$, where $a_{\rm e} = (4\pi n_{\rm e}/3)^{-1/3}$.     
Curves, describing the solid, contain nothing other than 
the electrostatic energies and residual entropies.

In the liquid, the free energies are given by the linear mixing rule based
on the classic fit \cite{PC00}
plus the correction \cite{P+09} and minus thermal free energies
of the solid state given by the linear mixing rule and comprised 
of classic harmonic terms 
(including the average phonon frequency logarithm) and classic anharmonic
terms of the first three orders \cite{BC22}. Since these contributions are independent of 
or linear in $x$, their subtraction
from the liquid quantities (instead of addition to the solid quantities) does not
alter the outcome of the double tangent construction. At $x=0$, the liquid free energy
thus defined becomes zero at temperature $T=T_{\rm 1m}$, where
$T_{\rm 1m} \equiv U_1/\Gamma_{\rm 1m}$ and $\Gamma_{\rm 1m}=175.719$.  
In Fig.\ \ref{dfs}, we plot the liquid free energy (minus the solid terms) at $T=T_{\rm 1m}$ 
by the black dashed curve.  

Open symbols in Fig.\ \ref{dfs} show numerical calculations 
of corrections to linear-mixing energies
of the solid state \cite{O+93} with circles representing the
``minimum'' 
Monte Carlo (MC)
energies, and squares corresponding to the fully-ordered
configurations, CsCl and $4\{{\rm fcc}\}$. Dotted and thick solid
curves show the fit proposed in \cite{O+93} and $U_{\rm rnd}$,
respectively.   
The latter provides a better fit to the open circles at $x=0.25$, 0.5, and
0.75 but a worse fit at $x=0.01157$ (cf. inset).
Such an agreement between first-principle and semi-analytic studies is 
extremely valuable. While the results of the former are reliable,
they provide little insight into the physical processes, occurring in the system.
By contrast, the results of the latter have clear physical meaning.

Dot-dashed and solid curves depict Helmholtz free energies. They are obtained 
from the dotted and thick solid curves (of the same color) by subtracting $TS_{\rm mix}/N$.  
Straight short-dashed grey line is the double-tangent construction for the 
random-solid model at $T=T_{\rm 1m}$. 

Let us note in passing that the horizontal line $F=0$ in Fig.\ \ref{dfs}
represents the free energy of a mechanical mixture of pure C and O
crystallites, whereas the open squares represent the free energies of
the respective ordered mixtures. Both types of systems have 
zero residual entropies.

\begin{figure}
\begin{center}
\includegraphics[bb=29 25 611 557, width=88mm]{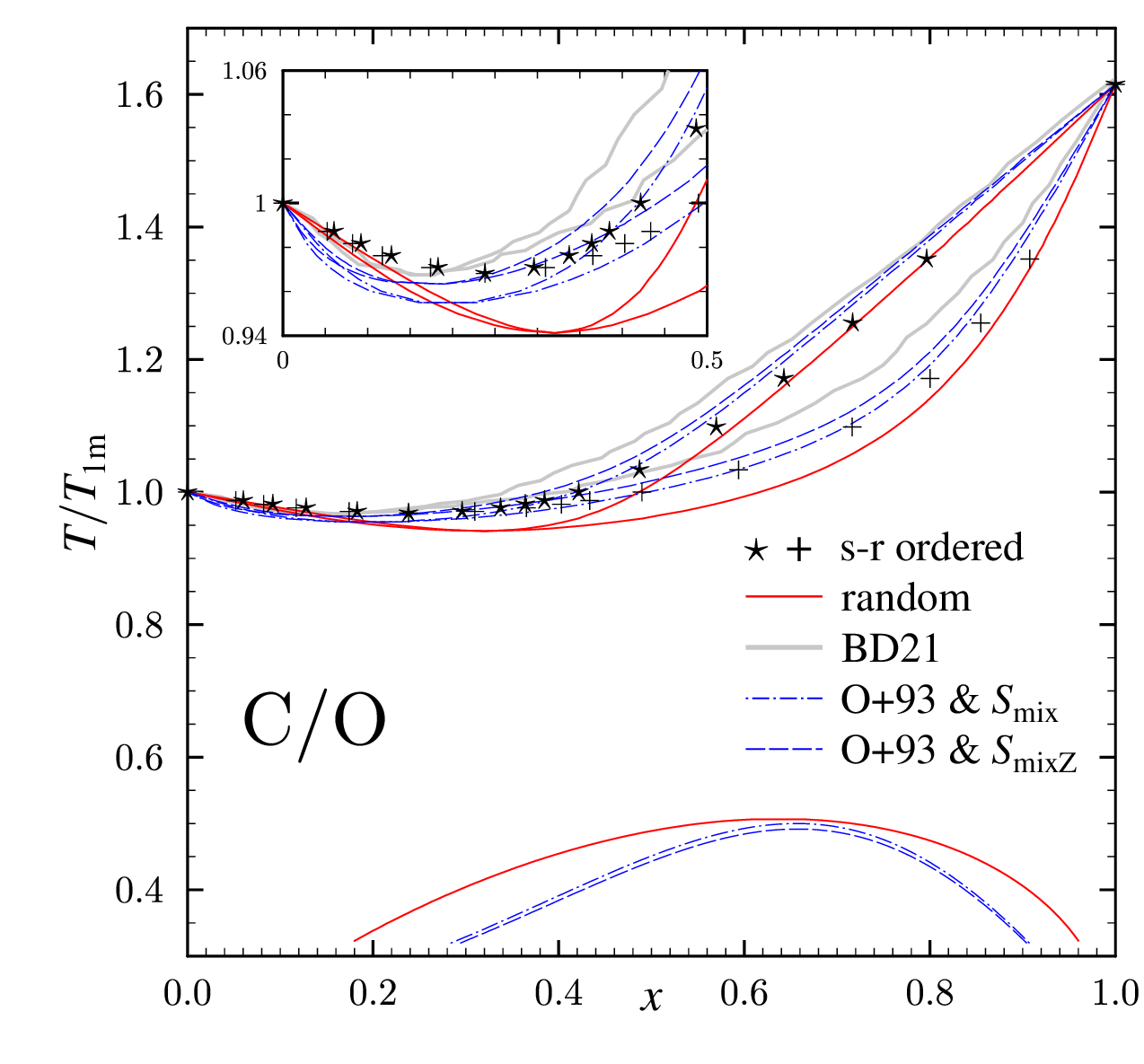}
\end{center}
\vspace{-0.2cm}
\caption[ ]{C/O phase diagram for a short-range ordered crystal (symbols),
fully disordered crystal (solid), based on simulations \cite{BD21} (thick solid grey),
based on the energy model \cite{O+93} with the residual entropy equal to $S_{\rm mix}$ (dot-dashed)
or $S_{\rm mixZ}$ (dashed).} 
\label{phased_rnd}
\end{figure}

Phase diagrams based on the free energy models, depicted 
in Fig.\ \ref{dfs},
are plotted in Fig.\ \ref{phased_rnd} by dot-dashed and solid lines for
fit \cite{O+93} and disordered crystal model, respectively. 
In the case of fit \cite{O+93},
by dashes, we also show the phase diagram for which the solid free 
energy is obtained from the energy by subtracting $TS_{\rm mixZ}/N$. The diagrams, originating from
fit \cite{O+93}, are the same as those obtained in \cite{B22}, and, in the
case of $S_{\rm mixZ}$ usage, the liquid-solid ``banana'' coincides with that 
of Ref.\ \cite{MC10}.      

{\it Short-range ordered C/O mixture} -- The question that naturally 
arises is whether a short-range order
in a mixture could result in a thermodynamically preferred state. To see 
this, one has to minimize the free energy with respect to the order
parameters. The latter must satisfy constraint (\ref{sum_rule}), inequality
above it,  
and the condition $\left| \Dpk \right|^2 \geq 0$
throughout ${\rm B}_1$.
In this work, we do not study oscillations of 
ions around their equilibrium positions. In principle,
another constraint should be that the frequencies of these oscillations
are real at any $\bm{k}$ in ${\rm B}_1$. 

\begin{figure}
\begin{center}
\includegraphics[bb=16 20 600 564, width=88mm]{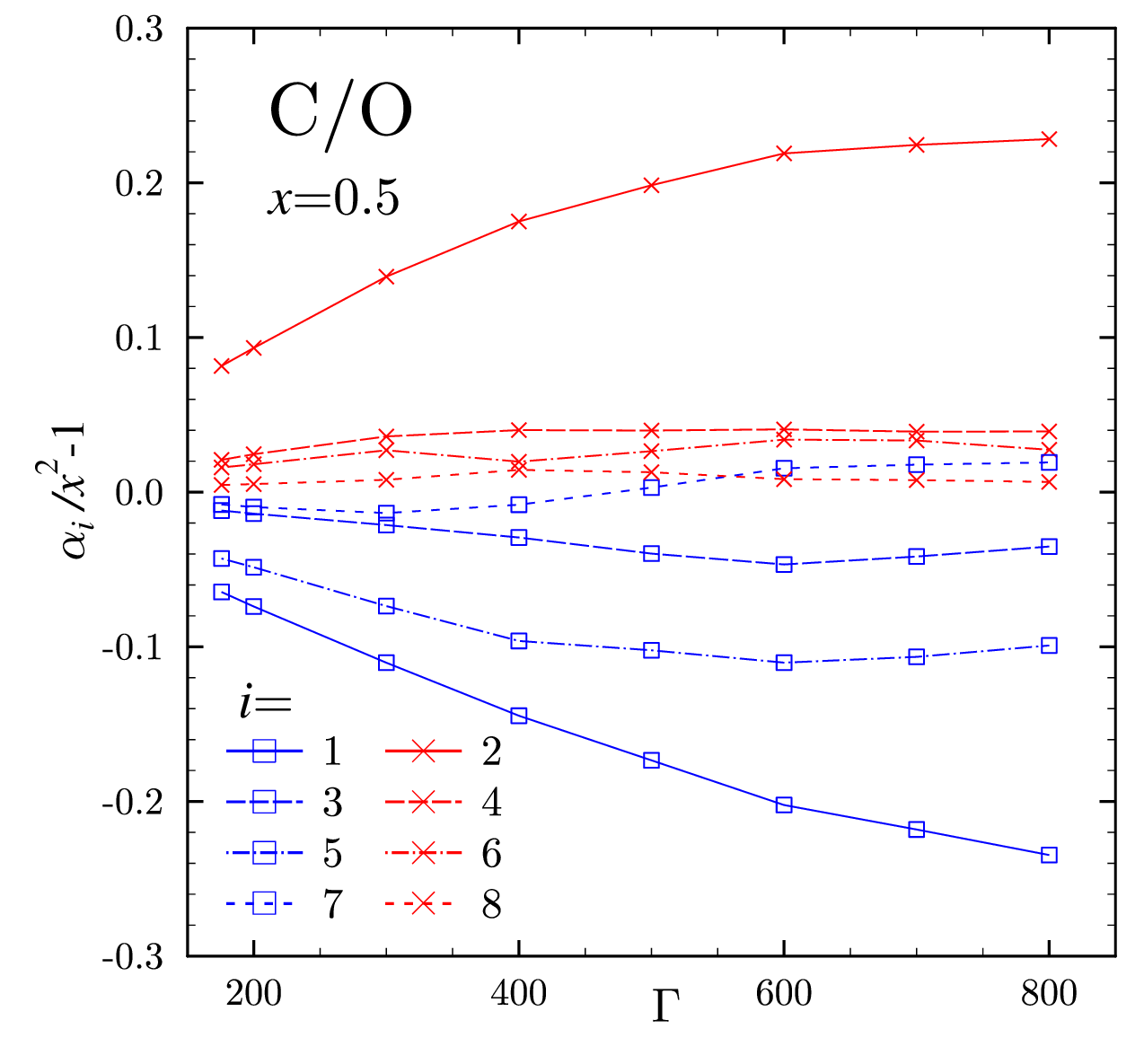}
\end{center}
\vspace{-0.2cm}
\caption[ ]{Odd (squares) and even (crosses) order parameters at free energy minimum.} 
\label{alpha_min}
\end{figure}

We have performed such minimization with 8 order parameters 
at various $x$ and $\Gamma$ values ($\Gamma \equiv U_1/T)$. 
In the process, we have checked the condition 
$\left| \Dpk \right|^2 \geq 0$ at $\sim 10^5$ wavevectors in the
primitive part of ${\rm B}_1$, including 600 points on its 6 edges 
(including vertices), 4000 random points on its 4 faces, and $\sim 10^5$ random
points in the inner region. Free energies minimized
with 7 or 8 order parameters differred very little, by 1-2 in the fourth significant
digit. Minimization with 6 order parameters was slightly worse at the 
highest considered $\Gamma \sim 700$-800.

In Fig.\ \ref{alpha_min}, we plot relative deviations from $x^2$ of 
the order parameters, realizing free energy minima, 
vs.\ $\Gamma$ at $x=0.5$.
Squares and crosses depict odd and even neighbor shells, respectively.
The lowest $\Gamma$ point is $\Gamma_{\rm 1m}$.

In Fig.\ \ref{dfs}, by filled dots, we show minimized crystal free energies
at $T=T_{\rm 1m}$.
A thin solid line,
going through the dots, is a segment of a natural cubic spline. Straight
long-dashed grey line is a double tangent between the spline and the liquid free energy. 

By repeating this construction at various values of $\Gamma$
from 130 to 181.5, we have mapped out the phase diagram of C/O liquid 
crystallization into a short-range ordered solid. It is shown by symbols in
Fig.\ \ref{phased_rnd}, where stars and pluses represent liquidus and solidus,
respectively. We note that the free energies of the short-range ordered solid
tend to free energies of the fully disordered solid at $x,1-x \ll 1$. As a result, 
symbols approach the solid curves in Fig.\ \ref{phased_rnd} at low and high
$x$. 

The models of fully disordered and short-range ordered solids
reported in this work are the first models of crystallized mixtures, 
not involving ab initio simulations, which predict azeotropic 
C/O phase diagram, thereby explaining the diagram shape in a transparent way.  
Thick solid grey curves in Fig.\ \ref{phased_rnd}
show currently the most advanced first-principle based C/O phase diagram \cite{BD21}.
We observe a decent agreement of our symbols with this diagram 
especially
at $x \lesssim 0.2$, 
indicating a preference for the short-range order. 
The agreement worsens at higher $x$, 
and, at this point, it is worth listing 
certain shortcomings of our present approach, which may be responsible
for the discrepancy. Firstly, our energy expression does not contain
cubic and higher-order terms in $\delta Z$, which would otherwise contribute to $U_{\rm shft}$.
Secondly, a more self-consistent expression for the residual entropy is desired.
Thirdly, the treatment of harmonic and anharmonic ion vibrations around
their equilibrium positions in the lattice is presently limited to the linear mixing rule
and can be improved. Such future improvement should also allow one to exclude phonon-unstable
configurations from the free energy minimization procedure.

\begin{figure}
\begin{center}
\includegraphics[bb=37 22 614 564, width=88mm]{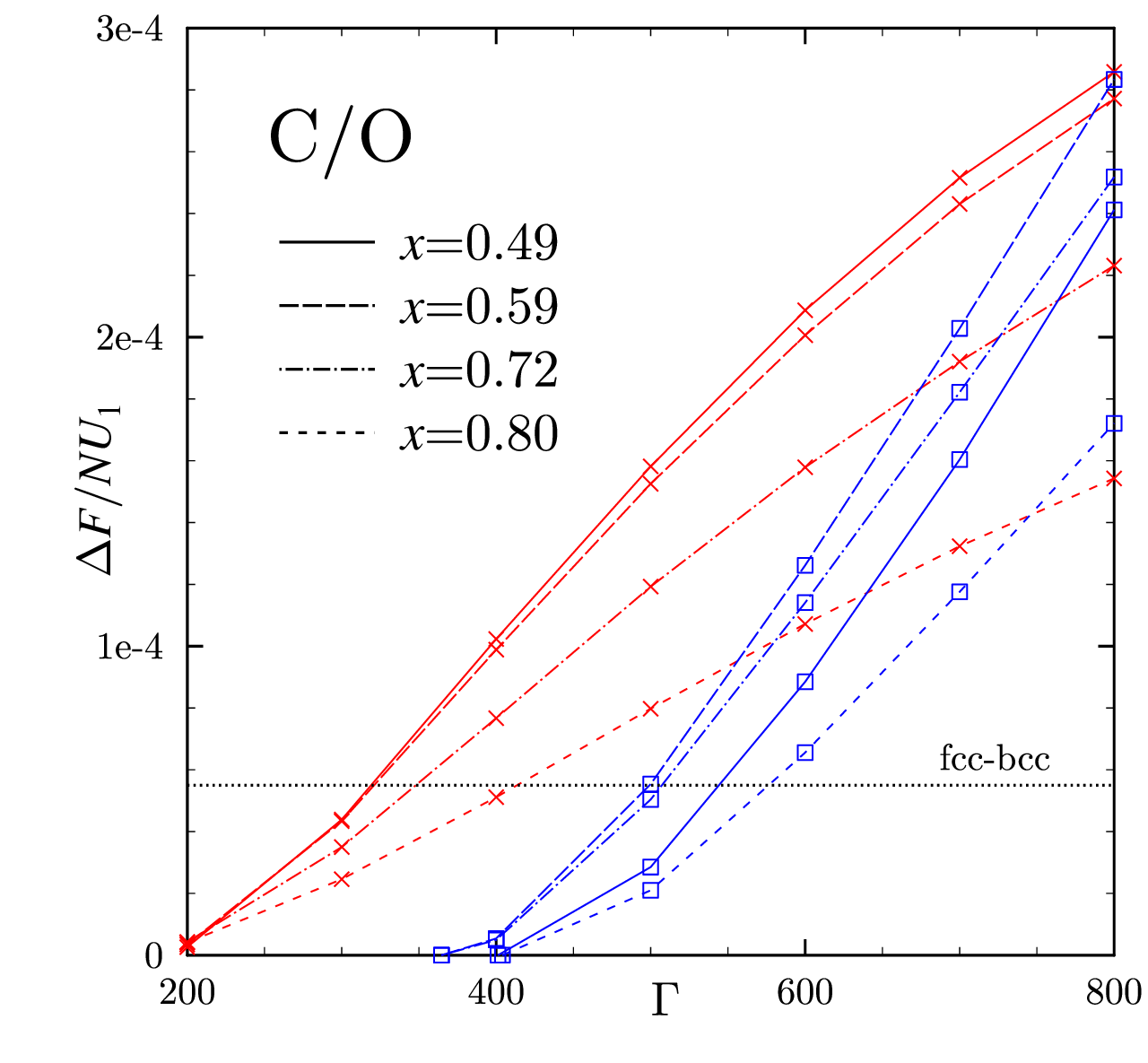}
\end{center}
\vspace{-0.2cm}
\caption[ ]{Free energy gain from short-range ordering (crosses), 
exsolution (squares), fcc to bcc structural transition (dotted).} 
\label{delta_f}
\end{figure}

{\it Equilibrium C/O crystal structure} -- Fig.\ \ref{alpha_min} indicates 
that the order parameters at
thermal equilibrium evolve with 
temperature. This means that whatever structure is formed upon
crystallization, it ceases to be a preferred state as soon
as the temperature drops. Free energy of this structure at a lower 
temperature can be obtained from 
its free energy at crystallization by rescaling the entropy term with 
the ratio of new and crystallization temperatures.
In Fig.\ \ref{delta_f}, we show by crosses the free energy
difference between such rescaled free energy 
and the true equilibrium free energy at low $T$ 
and four values of $x \approx 0.49$, 0.59, 0.72, and 0.80,
corresponding to crystallization at $\Gamma=\Gamma_{\rm 1m}$, 170, 160, 
and 150, respectively. For comparison, the energy difference between
one-component fcc and bcc crystals is shown by dots. 

This picture of evolving equilibrium structure of a crystal mixture is different
from the one implied by the lower portion of Fig.\ \ref{phased_rnd}.
In the latter, it is assumed that the mixture below crystallization
has a fixed microstructure but, at some critical $T$ and $x$, its free energy
becomes larger than the free energy of 
a mechanical sum of two such mixtures,
with lower and higher $Z_2$ fractions. This signifies the exsolution onset and
the appearance of the miscibility gap bounded by the solvus curve \cite{B22}.
The free energy gain, resulting from the exsolution process (for the dashed solvus
in Fig.\ \ref{phased_rnd}), is also shown in Fig.\ \ref{delta_f} by squares 
for the same values of $x$. It is typically smaller 
than the gain from crystal short-range reordering and, unlike the latter, it becomes 
nonzero not at the crystallization temperature but at the solvus temperature.      

It is hard to say what the fate of a crystal mixture, having a nonequlibrium
structure, is in actual degenerate stars. It is a question for kinetics.
We can assert though that if any kind of restructuring does occur with temperature decrease,
it will be accompanied by thermal energy release and an
effect on stellar cooling similar to that studied in \cite{B23,C+24}.

{\it Conclusion} -- We have proposed a new method of analysis of a crystallized binary
mixture of atomic nuclei on neutralizing electron background which 
serves as a model of matter in the inner layers of degenerate stars.
The method includes (i) introduction of short-range order parameters to
specify average mutual arrangement of nuclei of two sorts; (ii) exact 
expression for the electrostatic energy of a short-range ordered mixture,
assuming all ions are located precisely at the lattice nodes; 
(iii) lowest order correction to the electrostatic energy due to 
static ion displacements from their nodes;    
(iv) estimate of the mixture residual entropy with account of partial ordering;
(v) minimization of the free energy with respect to the order parameters
to determine the equilibrium state.
This is the first application of the short-range 
ordering idea to crystal mixtures of atomic nuclei and the first 
qualitative and quantitative description of the microstructure of 
these systems.

Being applied to a C/O mixture, the method allows one to (i) explain
the results of simulations \cite{O+93} with regard to ``minimum'' MC crystal
mixture energy; (ii) prove that C and O positions on a lattice are not 
random but are short-range ordered; (iii) find the order parameter dependence
on temperature; (iv) obtain C/O phase diagram in a decent agreement with the diagram
\cite{BD21}, the most recent one based on first-principle
simulations, especially in the region $x \lesssim 0.2$, where it is hard 
to achieve good resolution (cf.\ \cite{HSB10}); (v) reappraise the situation
below crystallization temperature,       
where instead of an abrupt transition to a state with a miscibility gap, 
the equilibrium state is predicted to have a continuously 
evolving structure.   
In summary, this is the first model, not relying on 
ab initio simulations of crystallized mixtures, 
which predicts azeotropic C/O phase diagram shape. 
By proposing, for the first time, a plausible ordering in the system,
the diagram is brought into a reasonable agreement with 
that obtained in the most recent first-principle study.  

In the future, the mixture microstructure obtained here can be a starting point 
for more detailed studies of ion thermodynamics
as well as for calculations of electron-ion scattering rates,
which determine kinetic coefficients.
The formalism can be readily extended to mixtures of more than two ion types
which, among other applications, would allow one to verify independently
$^{22}$Ne-depletion and buoyancy of crystals in C/O/$^{22}$Ne mixtures
\cite{I+91,BDS21} which is presently invoked for explanation of both
multi-Gyr WD cooling delays \cite{BBC24} and their strong magnetic field
generation \cite{L+24}.            

{\it Aknowledgment} -- The author is grateful to A.\ A.\ Kozhberov with whom the definition of 
$\delta Z(\bm{R})$ and its Fourier transform was discussed. 
This work was supported by Russian Science Foundation, grant 24-12-00320.

\appendix* 
\section{} 
{\it Residual entropy} -- Let us select
a vector $\bm{R}_{i{\rm n}}$ from an ion to its $i$th order neighbor.
There are $N$ such vectors in the lattice, $Nx$ of which originate from $Z_2$
while $N(1-x)$ from $Z_1$. Of the $Nx$ vectors, originating from $Z_2$,
a fraction $\alpha_i/x$ must point to another $Z_2$. Therefore, there 
are in total $N\alpha_i$ $[Z_2 Z_2]$ $\bm{R}_{i{\rm n}}$-links and $N(x-\alpha_i)$
$[Z_2 Z_1]$ $\bm{R}_{i{\rm n}}$-links. Thus, there are also $N(x-\alpha_i)$ $[Z_2 Z_1]$ 
links, corresponding to the opposite vector $-\bm{R}_{i{\rm n}}$.        
And thus there are in total $N(x-\alpha_i)$ and
$N(1-2x+\alpha_i)$ $[Z_1 Z_2]$ and $[Z_1 Z_1]$ $\bm{R}_{i{\rm n}}$-links, respectively.

We aim to express the entropy via these numbers of links,
which guarantee that the order parameters have the desired values.
Since the direct calculation is known to diverge \cite{C65}, we shall
consider the entropy decrement due to the difference between $\alpha_i$
and $x^2$. If all $\alpha_i=x^2$, then the total entropy is of
course $S_{\rm mix}$.

The calculation is based on the number of ways $N$ links can be
split into 4 groups such that permutations within each group do not matter.
Then the entropy decrement becomes
\beq
      \Delta S(\bm{R}_{i{\rm n}}) = \ln{\frac{H(\alpha_i)}{H(x^2)}}~,
\label{dSdef}
\eeq
where
\beq
    H(\alpha) = \frac{N!}{(N\alpha)! \{[N(x-\alpha))]!\}^2 [N(1-2x+\alpha)]!}~.
\label{Hdef}
\eeq
Using Stirling's formula, we obtain
\bea
   \frac{\Delta S(\bm{R}_{i{\rm n}})}{N} &=& 2 x\ln{x} + 2 (1-x) \ln{(1-x)} 
\nonumber \\
   &-& \alpha_i \ln{\alpha_i} 
   - 2(x-\alpha_i) \ln{(x-\alpha_i)} 
\nonumber \\
   &-& (1-2x+\alpha_i) \ln{(1-2x+\alpha_i)}~.      
\eea
Since there is nothing special about the vector $\bm{R}_{i{\rm n}}$, 
the total entropy decrement becomes 
\beq
   S-S_{\rm mix} = \frac{1}{2} \sum_{i} m_i \Delta S(\bm{R}_{i{\rm n}})~,
\label{Sres}
\eeq
where $1/2$ takes into account the fact that links, corresponding to
$\bm{R}_{i{\rm n}}$ and $-\bm{R}_{i{\rm n}}$ are the same.

Just like \cite{C65}, we realize that there are interactions between links,
corresponding to different vectors, which, presumably, will give rise
to contributions to equation (\ref{Sres}), containing higher powers of 
the order parameter. However, this problem is too complicated, and, for the time being, 
we treat different contributions as independent. 

The entropy $S$ given by equation (\ref{Sres}) is not bounded from 
below and, in particular, if $\alpha_i = x$ is assumed for all $i$, which corresponds
to separation of charges $Z_1$ and $Z_2$ and thus to zero entropy,
equation (\ref{Sres}) predicts $-\infty$. In our numerical work,
we have implemented a check of the entropy sign and did not allow states with $S<0$.      
However, this has not affected results reported in this paper 
because all configurations, realizing free energy minima,
had residual entropies well above zero.

\end{document}